\documentclass[letterpaper]{article}
\usepackage{flairs}
\usepackage{times}
\usepackage{helvet}
\usepackage{courier}
\usepackage[hyphens]{url}
\usepackage{hyperref}
\usepackage{graphicx}
\graphicspath{ {figures/} }
\usepackage{caption}
\usepackage{subcaption}
\usepackage{wrapfig}
\usepackage{multirow}
\usepackage{tabularx}
\usepackage{paralist}
\usepackage{colortbl}
\usepackage{dingbat, pifont}
\usepackage{listings}
\usepackage[T1]{fontenc}
\usepackage{tikz}
\usepackage{float}
\usepackage{amsmath}
\usetikzlibrary{arrows.meta, shapes, positioning, calc}
\definecolor{Gray}{gray}{0.9}

\usepackage{array} 

\begin{document}
%
\title{Towards Deterministic Sub-0.5 µs Response on Linux through Interrupt Isolation}
\author{
\begin{tabular}{llllll}
  Zhouyi Zhou & Zhili Liu & Shancong Zhang & Jiemin Li & Dengke Du & Mengke Sun \\
  Zhiqiang Wang & Hongyan Liu & Guokai Xu \\
\end{tabular}
\\
Beijing Ucas Technology Co,Ltd., Beijing, China\\
\{zhouyi.zhou, zhili.liu, sam.zhang, jiemin.li,
dengke.du, mengke.sun,\\ zhiqiangwang, hongyan.liu, xuguokai\}@ucas.com.cn\\
}
\maketitle

\begin{abstract}
Real-time responsiveness in Linux is often constrained by interrupt contention and timer
handling overhead, making it challenging to achieve sub-microsecond latency.
This work introduces an interrupt isolation approach that centralizes and minimizes timer
interrupt interference across CPU cores. By enabling a dedicated API to selectively invoke
timer handling routines and suppress non-critical inter-processor interrupts,
our design significantly reduces jitter and response latency. Experiments conducted on an ARM-based multicore
platform demonstrate that the proposed mechanism consistently achieves sub-0.5 µs response
times, outperforming conventional Linux PREEMPT-RT configurations. These results highlight
the potential of interrupt isolation as a lightweight and effective strategy for
deterministic real-time workloads in general-purpose operating systems.
\end{abstract}

\section{Introduction}

Modern high-performance and real-time systems increasingly demand 
\textbf{predictable CPU behavior}, \textbf{ultra-low interrupt latency}, 
and \textbf{minimal system noise}. 
However, general-purpose operating systems such as Linux were not 
originally designed to meet such stringent timing requirements. 
As a result, applications with tight latency constraints often suffer from 
\textbf{execution disturbances} caused primarily by two kernel-level 
mechanisms: \emph{periodic timer interrupts} and 
\emph{inter-processor interrupts (IPIs)}.  

\subsection{Timer Interrupts}
In Linux, periodic timer interrupts (commonly referred to as \emph{ticks})
are scheduled on every CPU core to maintain system time and perform
kernel housekeeping tasks\cite{Corbet}. This design reflects the traditional reliance
of Unix-like operating systems on timer interrupts for core kernel
functions, as documented in 
``Clock Interrupt Handling''~\cite[Section~5.2]{Vahalia96}.
These interrupts invoke routines for timekeeping, task scheduling,
timer expiration, bandwidth enforcement for real-time processes,
and other internal kernel mechanisms. While harmless---or even
beneficial---for general workloads, such periodic processing can
be highly disruptive for latency-sensitive applications such as
user-space packet processing, real-time control loops, and multimedia
pipelines, where \textbf{deterministic CPU execution} is critical.

Mechanisms such as \texttt{NOHZ\_FULL} attempt to mitigate this problem
by reducing tick frequency, but they cannot fully eliminate timer activity.
Even in \texttt{nohz\_full} mode, periodic ticks still occur, leaving
``isolated'' cores subject to residual interruptions.

\subsection{Inter-Processor Interrupts (IPIs)}
In addition to timer ticks, IPIs represent another significant source of 
jitter. Within the Linux kernel, IPIs are widely used to coordinate state 
changes across cores---for example, to initiate TLB shootdowns, 
wake remote tasks, or synchronize RCU grace periods. 
From the perspective of an application pinned to a dedicated core, 
IPIs are asynchronous and unpredictable, often leading to unwanted 
preemptions or context switches.  

Disabling local timer interrupts reduces some noise but does not prevent 
a core from receiving IPIs initiated elsewhere in the system. 
Consequently, even cores exempt from ticks may still face arbitrary 
interruptions caused by global kernel coordination mechanisms.  

\subsection{Motivation}
These limitations highlight a fundamental challenge:
\textbf{conventional mechanisms are insufficient for achieving sub-microsecond, deterministic response times in Linux}.
In latency-critical domains such as high-frequency trading, average latency has already been driven to the nanosecond scale, making \textbf{jitter—even at nanosecond resolution—the decisive competitive factor}  \cite{risk2024ultra}.  Systems that sustain nanosecond-level determinism consistently outperform those with microsecond variability.
To address this, we propose an \textbf{interrupt isolation framework} that centralizes timer interrupt handling and suppresses unnecessary IPIs, enabling Linux systems to deliver \textbf{sub-0.5~$\mu$s response times} for real-time workloads.

\subsection{Our Approach: Full Isolation via Centralized Timer Handling and Shared-Memory IPI Management}

In this paper, we present a novel and more aggressive approach to CPU isolation in Linux, 
designed to achieve deterministic sub-microsecond response times. 
Our mechanism introduces two key innovations:

\begin{itemize}
\item \textbf{Centralized Timer Handling:} Centralized Timer Handling. We centralize tick work that previously executed on the isolated core
  and expose it through a dedicated polling API. This API is not automatic; the real-time user code must call it periodically
  (or at application-chosen safe points). If the tick work is completely disabled—so that functions such as \texttt{account\_process\_tick} are
  never invoked—these routines simply do not execute. While this does not compromise system stability, it suppresses time accounting and related statistics.
  Users who require these updates can invoke the API on demand to run the deferred tick work.
    
    \item \textbf{Shared-Memory IPI Management:} Non-essential inter-processor interrupts (IPIs) to isolated cores 
    are suppressed. Coordination that would normally trigger IPIs is instead implemented via shared memory, 
    allowing controlled and predictable communication without asynchronous preemption.

    The suppression mechanism works by:
    
    1. Masking actual IPI delivery to isolated cores
    
    2. Providing virtual IPI completion status to the sender
    
    3. Ensuring synchronization primitives (e.g., TLB shootdowns) appear to complete
    
\end{itemize}

This level of isolation is not achievable with existing mechanisms such as 
\texttt{NOHZ\_FULL} or \texttt{isolcpus} alone. 
By explicitly separating time-related kernel responsibilities and reassigning them to dedicated API, 
and by handling inter-core coordination through shared memory rather than IPIs, 
isolated cores execute continuously without interruption.  

The proposed approach is particularly effective in latency-critical scenarios, 
such as GPIO toggling, real-time control loops, and user-space packet processing. 
It allows the Linux kernel to efficiently manage global interrupts, 
isolate real-time workloads on dedicated cores, and guarantee timely execution 
of high-precision tasks free from interference by timer events or other system operations.

\section{Related Work}\label{BG}
\subsection{Timer Interrupt Control in the Linux Kernel}

The concept of timer ticks in operating systems predates the development of the Linux kernel \cite{Corbet}. 
In Linux, periodic timer interrupts, commonly referred to as \emph{ticks}, are scheduled on each CPU core 
to support timekeeping, process accounting, and scheduler activation. These interrupts typically occur at 
frequencies between 250 Hz and 1 kHz and may preempt running tasks to perform kernel maintenance.  

To reduce the overhead caused by frequent ticks, \emph{tickless} systems were introduced. 
The \texttt{NO\_HZ} feature disables timer interrupts on idle CPUs, and its extension, 
\texttt{NOHZ\_FULL}, attempts to eliminate ticks even during active user-space execution \cite{KernelDocNOHZ}. 
However, timer ticks still occur on \texttt{NO\_HZ}/\texttt{NOHZ\_FULL} CPUs because the kernel must perform 
statistics collection and other internal bookkeeping \cite{stackoverflow} \footnote{ Upon further examination of
the latest kernel code \cite{torvalds2025linux}, the function tick\_nohz\_idle\_stop\_tick which stops
the timer ticks is only invoked when the system enters an idle state. In our specific application
scenario, isolated CPU cores will not enter an idle state as long as they are running applications.}.

While these mechanisms help reduce timer-induced jitter, they do not completely eliminate all sources of latency, 
leaving real-time workloads exposed to residual interruptions.

\subsection{Inter-Processor Interrupts (IPIs) and Their Effects}

Inter-processor interrupts (IPIs) in Linux can significantly impact real-time performance by introducing 
unpredictable latency. IPIs force a CPU to handle interrupts originating from another processor, 
potentially preempting time-critical tasks. In real-time systems, where deterministic scheduling and 
minimal latency are essential, IPIs—used for operations such as TLB flushes, cache management, 
and scheduler balancing—can cause delays if triggered during the execution of high-priority tasks. 
Unlike timer interrupts, IPIs are externally triggered and often asynchronous. Consequently, even 
tickless CPUs may receive unexpected IPIs, introducing nondeterministic latency and potential cache 
disruptions. Existing research has identified IPIs as a major obstacle to achieving strict real-time 
performance on Linux \cite{YuxinRen}.  

To mitigate the impact of IPIs on real-time workloads, it is necessary to systematically identify and 
address each source of cross-core interrupts. For example, unloading block devices in Linux triggers 
IPIs, so such operations should be avoided in latency-critical environments. Similarly, CPU frequency 
statistics require IPIs to query other cores; disabling dynamic frequency scaling or using a fixed-frequency 
mode eliminates this source of latency. By analyzing and selectively suppressing IPI-inducing operations—such as 
TLB shootdowns, RCU callbacks\footnote{RCU uses IPIs for expedited grace periods, but we disable expedited
grace periods for real-time workloads by using rcupdate.rcu\_normal},
and scheduler balancing—we can remove unnecessary cross-core interruptions, 
ensuring deterministic execution for real-time tasks. Combined with CPU isolation and IRQ affinity tuning, 
this targeted approach helps maintain hard real-time guarantees on Linux systems.

\subsection{CPU Isolation Techniques}

Linux provides mechanisms such as \texttt{isolcpus} and IRQ affinity \cite{KernelDocParam} 
to isolate processor cores from the general scheduler and external interrupts. 
These tools can effectively prevent a core from executing most kernel tasks and services, 
including RCU callbacks. However, they do not block all types of interruptions. 
In particular, timer interrupts and inter-processor interrupts (IPIs) may still be delivered 
to an isolated core. Consequently, while \texttt{isolcpus} and IRQ affinity reduce interference 
from the operating system, they do not achieve complete isolation at the hardware interrupt level.  

Redhawk Linux RTOS \cite{redhawk} further minimizes the impact of clock interrupts on real-time workloads 
through specialized kernel modifications. These mechanisms significantly reduce the influence 
of inter-core interrupts on system performance. Nonetheless, Redhawk does not entirely eliminate 
inter-core interrupts on isolated cores; instead, they may still occur conditionally under certain scenarios.

\section{Methodology}\label{BE}

This section describes the kernel-level design, isolation strategy, and implementation workflow we employed
to eliminate system noise on real-time application cores without delegating maintenance tasks to external
housekeeping cores. In contrast to traditional separation approaches, our system uses a novel method
where the application core invokes clock maintenance functions on demand via a centralized interface.
This design is made possible through the custom Isolator API v0.1, which we developed to enable complete
tick and IPI suppression while preserving system functionality through cooperative execution.

\subsection{System Design Overview}
\label{subsec:system-design}

The system architecture is designed to provide a robust, low-latency environment for real-time applications by leveraging the concept of core isolation. The central premise is to dedicate one or more processor cores to run real-time tasks exclusively, shielding them from non-deterministic interference from the general-purpose operating system (OS), specifically Linux. This is achieved through a managed isolation lifecycle that suppresses OS services and then reintroduces them in a controlled manner for necessary maintenance.

\begin{figure}[tb]
    \centering
    \includegraphics[width=0.9\linewidth]{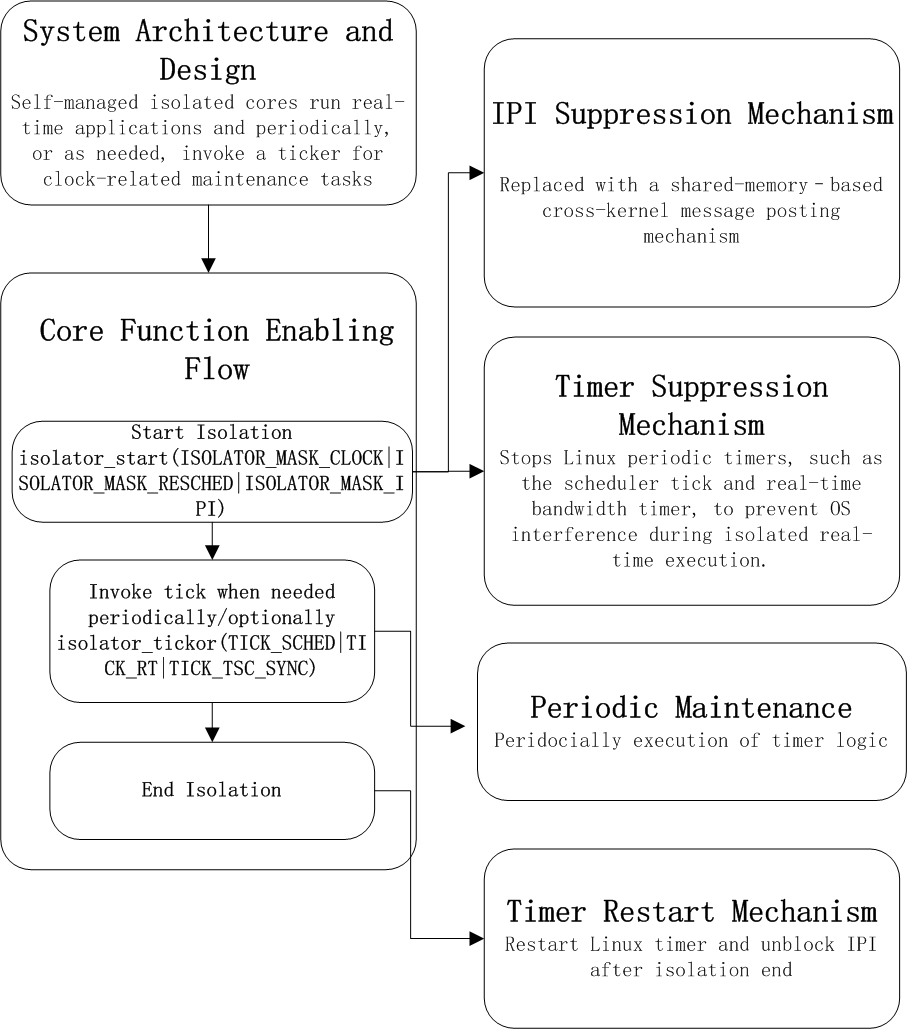}
    \caption{Core isolation enabling flow and mechanisms.}
    \label{fig:isolator_flow}
\end{figure}

The design, illustrated in Figure~\ref{fig:isolator_flow}, is built around three core mechanisms:

\begin{enumerate}
    \item \textbf{Isolation Enabling:} The process begins by invoking \texttt{isolator\_start} with a specified mask (e.g., \texttt{ISOLATOR\_MASK\_CLOCK}, \texttt{ISOLATOR\_MASK\_RESCHED}, \texttt{ISOLATOR\_MASK\_IPI}). This function initiates the isolation state on a core, effectively severing its interaction with standard OS interrupt mechanisms. Crucially, this involves suppressing Inter-Processor Interrupts (IPIs) by replacing them with a deterministic, shared-memory-based message-passing system for cross-kernel communication. It also halts Linux's periodic timers, such as the scheduler tick and real-time bandwidth timer.

    \item \textbf{Periodic Maintenance via a Tick:} While isolated, the core executes its real-time workload without interruption. However, to maintain system health and coherence, certain clock-related functions must still be performed periodically or on-demand. This is handled by the \texttt{isolator\_tick} function, which is invoked by the application itself. This function acts as a managed tick, allowing the isolated core to voluntarily execute maintenance tasks---such as scheduling decisions (\texttt{TICK\_SCHED}), real-time task management (\texttt{TICK\_RT}), and time synchronization (\texttt{TICK\_TSC\_SYNC})---at a time of its choosing, thus preserving determinism. The basic idea of delegating periodic system duties to explicit user-invoked calls can be traced back to early real-time computing practices, as discussed in the first chapters of~\cite{mellichamp1983}.

    \item \textbf{Isolation Teardown:} Upon completion of the real-time task, the isolation state is ended. This final stage gracefully reintegrates the core back into the broader OS environment. The mechanism restarts the standard Linux periodic timers and re-enables the full IPI handling, ensuring the core returns to normal operation without any residual state from the isolation period.
\end{enumerate}

In summary, the design replaces the traditional preemptive, interrupt-driven OS model on isolated cores with a self-managed, cooperative paradigm. The real-time application retains full control over when and how maintenance occurs, resulting in significantly improved timing predictability and performance for latency-sensitive workloads.

\subsection{Interrupt Suppression and Timer Bypass}
\label{subsec:interrupt-timer}

The kernel was modified to explicitly recognize the isolation state of a core.
When a core is set to isolation mode, its local tick logic, normally executed in
\texttt{update\_process\_times} and \texttt{scheduler\_tick},
is completely bypassed unless explicitly triggered through the \texttt{isolator\_tick} function.
In addition, all asynchronous inter-processor interrupts (IPIs) that might interfere with the
isolated core are suppressed. The key mechanisms implemented for timer bypass and interrupt suppression are described below.

\paragraph{Timer Suppression}  
The \texttt{isolator\_start} function halts Linux’s periodic tick timer by calling \texttt{hrtimer\_cancel} to
prevent interference from the operating system. In addition, real-time bandwidth timers are maintained in a list,
and each active timer is iteratively canceled to stop any scheduled callbacks. As a result, both the scheduler tick
and other periodic kernel activities on the isolated core are effectively stopped.

\paragraph{IPI Suppression and Replacement}  
Asynchronous IPIs, which are normally used for inter-core notifications, are replaced with a deterministic,
shared-memory-based message posting mechanism. This enables controlled core-to-core communication without
the unpredictability of standard IPIs. Specific handling includes:
\begin{itemize}
    \item \textbf{RCU (Read-Copy Update):}  Upon entering isolation, the RCU subsystem is notified that the core is isolated and must be excluded from grace-period calculations. RCU already provides this behavior for CPUs running userspace code \cite{weisbecker2012context}. Our extension also covers cases where the real-time workload executes in kernel context, in such situations, excluding the isolated core from RCU observation must be handled carefully, since any kernel code that is not under RCU's watch cannot safely traverse RCU-protected data structures. 

    \item \textbf{TLB Shootdowns:} IPIs requesting Translation Lookaside Buffer flushes are suppressed and queued in shared memory.
      The isolated core processes these flush requests the next time it invokes \texttt{isolator\_tick}.
    \item \textbf{CPU Frequency Scaling:} IPIs induced by CPU frequency changes are disabled to avoid unwanted interruptions.
\end{itemize}

\paragraph{Periodic Maintenance}  
Deferred timer tasks and queued IPI processing are executed only when the isolated core voluntarily calls \texttt{isolator\_tick}. This ensures deterministic behavior while still allowing the core to perform necessary maintenance tasks at controlled intervals. \texttt{isolator\_tick} can be invoked from both user space and kernel space. Currently, in order to support invocation from user space, the entry from user space into kernel space is implemented via the proc filesystem.

\paragraph{Restoration Mechanism}  
When isolation ends, previously canceled Linux timers and real-time bandwidth timers are restarted, and normal IPI handling is restored. The core then returns to standard operation, with all periodic activities and inter-core notifications resumed.

\subsection{Memory and Communication Management}
\label{subsec:memory-communication}

Memory operations required during isolation are handled using a dedicated SLUB allocator reserved for the isolated core. All allocations and deallocations occur within this core-specific SLUB, and cross-core memory maintenance or access is strictly prohibited. This ensures that memory operations remain deterministic and fully isolated from other cores, avoiding interference from general-purpose kernel allocators.

For core-to-core interactions, a shared-memory-based messaging mechanism is employed to transfer messages between the isolated core and non-isolated cores. This mechanism eliminates the need for interrupts to wake remote threads, preserving the deterministic behavior of the isolated core. The primary communication methods are \texttt{post} and \texttt{poll}. Messages intended for a remote core are written into a shared memory region using \texttt{post}, while the receiving core periodically checks for new messages using \texttt{poll}. Callback functions can be registered to handle incoming messages in a non-intrusive, user-driven manner. This design guarantees that all communication occurs without generating asynchronous interrupts, allowing the isolated core to exchange information safely while maintaining strict determinism.

\subsection{Isolation Monitoring and Debugging}
\label{subsec:isolation-monitoring}

To support verification and monitoring of isolated cores, we implemented a real-time statistics collector. This collector tracks the types and counts of interrupts that are blocked or deferred on each core. By analyzing these statistics, developers can verify the effectiveness of the isolation mechanisms and assess the impact of different workloads and tuning configurations. This monitoring infrastructure provides a clear view of the core’s behavior under isolation, enabling precise debugging and performance evaluation.

\section{Implementation Details}
This section presents a detailed overview of the kernel-level modifications we introduced to
support high-determinism, low-jitter execution on isolated CPUs for high-performance, real-time
packet processing. These modifications are implemented within four critical subsystems of the
Linux kernel: RCU (Read-Copy-Update), the Real-Time Scheduler, SMP/IPI handling, and Timer Tick
Management. 

\subsection{Kernel Modifications Overview}
To enable CPU isolation at a fine granularity and ensure minimal OS interference on application-
dedicated cores, we performed targeted changes across the kernel. The key goals are to:
Eliminate unnecessary kernel activity on isolated CPUs,
Avoid inter-processor communication (e.g., IPIs) to reduce latency spikes,
Defer time-based maintenance (e.g., scheduling enforcement) to application-controlled contexts,
To coordinate the behavior of different kernel components, we define the following per-CPU data
structure:

\begin{figure}[H]
    \centering
    \begin{lstlisting}[language=C, caption={Per-CPU isolator counter definition}]
    DEFINE_PER_CPU(atomic_t, isolator_counters[CONFIG_NR_CPUS]);
    \end{lstlisting}
\end{figure}
Each entry in this array acts as a flag indicating whether the corresponding CPU is currently under
isolation. This flag is incremented by user-space applications via a custom system call or API
(isolator\_start) and is read by kernel components to selectively disable or defer their operations.

\subsection{RCU Grace Period Handling}
This modification ensures that CPUs marked as isolated are treated as being in a permanent
quiescent state by the RCU subsystem. By doing so, the system avoids unnecessary waiting on
isolated cores during RCU grace periods. This approach maintains forward progress of the RCU
mechanism while eliminating dependencies on cores that are no longer expected to participate in
normal kernel activity.

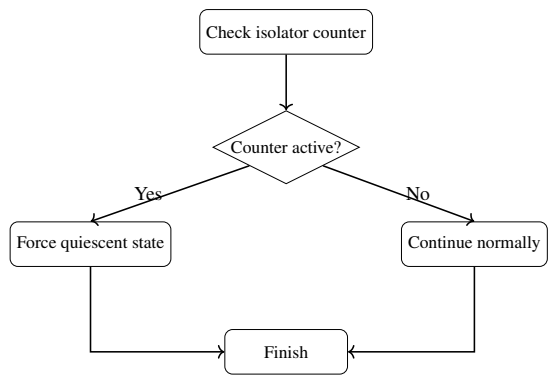
\begin{figure}[t]
\centering
\resizebox{0.85\columnwidth}{!}{%
\begin{tikzpicture}[node distance=1.0cm, auto]
  \tikzstyle{block} = [rectangle, draw, rounded corners,
                       minimum width=2.2cm, minimum height=0.8cm, 
                       align=center, font=\small]
  \tikzstyle{decision} = [diamond, draw, aspect=2, align=center, 
                          inner sep=1pt, font=\small]
  \tikzstyle{arrow} = [->, thick]

  \node[block] (start) {Check isolator counter};
  \node[decision, below=of start] (cond) {Counter active?};
  \node[block, below left=1.0cm and 1.4cm of cond] (force) {Force quiescent state};
  \node[block, below right=1.0cm and 1.4cm of cond] (normal) {Continue normally};
  \node[block, below=2.6cm of cond] (end) {Finish};

  \draw[arrow] (start) -- (cond);
  \draw[arrow] (cond.south west) -- (force.north) node[midway, left] {Yes};
  \draw[arrow] (cond.south east) -- (normal.north) node[midway, right] {No};
  \draw[arrow] (force.south) -- ++(0,-0.4) |- (end.west);
  \draw[arrow] (normal.south) -- ++(0,-0.4) |- (end.east);
\end{tikzpicture}%
}
\caption{Flowchart for enforcing quiescent state when the isolator counter is active.}
\label{fig:quiescent}
\end{figure}

Another important technical detail is that we explicitly prevent the \texttt{rcu\_preempt} kernel 
thread from running on isolated CPUs. During our experiments, we observed that the scheduler 
occasionally places the \texttt{rcu\_preempt} kernel thread on an isolated core. Since these cores 
are typically occupied by uninterrupted, high-priority user tasks, the \texttt{rcu\_preempt} 
thread may not get scheduled in a timely manner. This situation blocks the progress of RCU grace 
periods, preventing \texttt{synchronize\_rcu} from returning, which in turn causes other processes 
invoking \texttt{synchronize\_rcu} to remain blocked. By disallowing \texttt{rcu\_preempt} on 
isolated CPUs, we avoid this pathological scenario and preserve the expected non-blocking behavior 
of RCU for the rest of the system.

\subsection{Time Scheduler Deferral}
\paragraph{Tick Subsystem Modifications.}
The Linux \texttt{tick-sched} subsystem manages periodic scheduling ticks and 
nohz (tickless) operation. By default, each CPU receives periodic tick events 
driven by high-resolution timers, which are used to drive process accounting, 
timekeeping, and load balancing. However, on isolated CPUs dedicated to 
real-time packet processing, these periodic ticks represent an unnecessary 
source of nondeterministic interruptions.

To address this, we introduced a per-CPU flag (\texttt{ucas\_counter}) that 
controls whether tick-related hrtimers should be restarted or suppressed. When 
this counter is non-zero, functions such as \texttt{tick\_nohz\_restart}, 
\texttt{tick\_nohz\_idle\_restart\_tick}, \texttt{tick\_nohz\_handler}, 
and \texttt{tick\_sched\_timer} are modified to immediately return without 
restarting the periodic scheduler timer. This effectively disables the 
recurring tick on isolated CPUs, preventing spurious wakeups and involuntary 
kernel execution.

In order to preserve correctness, we provide helper routines that allow the 
isolator subsystem to explicitly handle deferred tick events when needed. For 
example, the function \texttt{ucas\_handle\_tick\_sched\_timer} invokes 
\texttt{tick\_sched\_do\_timer} and \texttt{update\_process\_times} 
manually, ensuring that accounting and timekeeping can still progress under 
controlled conditions.

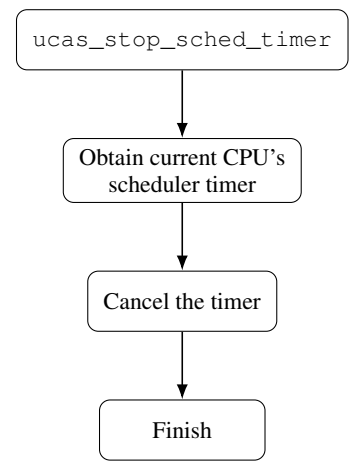
\begin{figure}[t]
\centering
\begin{tikzpicture}[node distance=0.9cm, auto]
  \tikzset{
    block/.style={
      rectangle, draw, rounded corners,
      minimum width=2.2cm, minimum height=0.8cm,
      inner xsep=6pt, inner ysep=4pt,
      align=center, font=\small
    },
    arrow/.style={-Latex, semithick}
  }

  \node[block] (start) {\texttt{ucas\_stop\_sched\_timer}};
  \node[block, below=of start] (get) {Obtain current CPU's\\scheduler timer};
  \node[block, below=of get] (cancel) {Cancel the timer};
  \node[block, below=of cancel] (end) {Finish};

  \draw[arrow] (start) -- (get);
  \draw[arrow] (get) -- (cancel);
  \draw[arrow] (cancel) -- (end);
\end{tikzpicture}
\caption{Flowchart for stopping the per-CPU scheduler timer.}
\label{fig:stoptick}
\end{figure}

We also provide utility functions such as \texttt{ucas\_stop\_sched\_timer} to stop
the tick-related hrtimers immediately Figure~\ref{fig:stoptick}.

This design guarantees that isolated CPUs are free from involuntary tick 
interrupts during real-time workloads, while still allowing deferred handling 
of system housekeeping activities when explicitly triggered. The result is a 
significant reduction in jitter and improved determinism for 
latency-sensitive applications.

\paragraph{Real-Time Bandwidth Deferral.}
In addition to suppressing periodic tick timers, we extended the real-time (RT) 
scheduler’s bandwidth control logic in \texttt{kernel/sched/rt.c}. In the 
unmodified kernel, each CPU uses a periodic high-resolution timer 
(\texttt{rt\_period\_timer}) to enforce CPU runtime quotas for SCHED\_RT tasks. 
While this mechanism ensures fairness and prevents runaway RT tasks, it also 
introduces recurring interrupts that can preempt high-priority user-level 
workloads running on isolated CPUs.

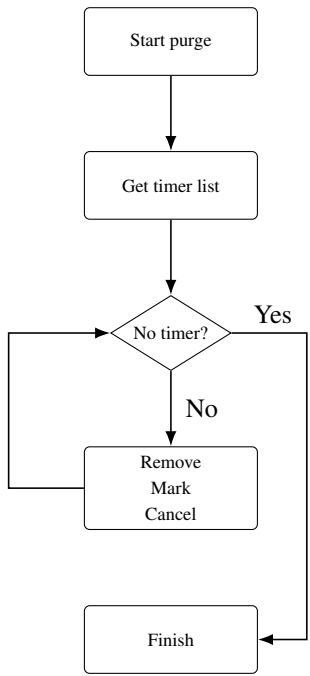
\begin{figure}[t]
\centering
\begin{tikzpicture}[
  node distance=1.0cm and 1.4cm,
  >=Latex,
  block/.style={
    rectangle, rounded corners=2pt, draw,
    align=center, font=\scriptsize,
    minimum width=2.3cm, minimum height=0.9cm
  },
  decision/.style={
    diamond, draw, aspect=1.4, align=center,
    font=\scriptsize, inner sep=1pt,
    minimum width=1.6cm, minimum height=1.0cm
  },
  arrow/.style={->, semithick}
]

\node[block]    (start)   {Start purge};
\node[block,    below=of start] (get) {Get timer list};
\node[decision, below=of get]   (cond) {No timer?};
\node[block,    below=of cond]  (proc) {Remove \\ Mark \\ Cancel};
\node[block,    below=of proc]  (finish) {Finish};

\draw[arrow] (start) -- (get);
\draw[arrow] (get) -- (cond);
\draw[arrow] (cond.south) -- node[right, xshift=2pt] {No} (proc.north);
\draw[arrow] (cond.east) -- ++(1.0,0) node[above, pos=0.55] {Yes} |- (finish.east);
\draw[arrow] (proc.west) -- ++(-1.0,0) |- (cond.west);

\end{tikzpicture}
\caption{Flowchart for purging per-CPU real-time bandwidth timers.}
\label{fig:purge_flowchart}
\end{figure}

To eliminate this source of interference, 
We provide utility functions such as 
\texttt{ucas\_purge\_this\_cpu\_rtb} to explicitly cancel real-time bandwidth timers when 
isolation is active. We augmented the \texttt{rt\_bandwidth} 
structure with a list node and a \texttt{deleted} flag. Each CPU maintains a 
per-CPU list of active bandwidth structures, \texttt{ucas\_purge\_this\_cpu\_rtb} cancel
the timers on above per-CPU list one by one Figure~\ref{fig:purge_flowchart}.

\paragraph{Summary.}
Together, the suppression of periodic tick timers and the deferral of real-time 
bandwidth enforcement create an execution environment where isolated CPUs are 
shielded from involuntary kernel activities. Tick-related housekeeping and 
scheduler quota enforcement are both deferred until explicitly invoked, ensuring 
that high-priority user workloads can proceed with minimal interference. 

\subsection{IPI Suppression}
The suppression of inter-processor interrupts (IPIs) to isolated CPUs eliminates one of the major
sources of external interference and cache coherence disruption in multicore systems. By
modifying the logic responsible for IPI target selection, the kernel dynamically filters out isolated
CPUs from cross-core signaling paths.
This ensures that CPUs dedicated to real-time or latency-sensitive processing remain unaffected
by kernel events originating from other cores, thereby improving predictability and scalability.

\subsection{Tickless Operation Enforcement}
The enforcement of tickless operation for isolated CPUs is achieved by disabling the kernel's
internal logic for restarting periodic scheduler ticks. Isolated CPUs are instead allowed to operate
independently of the kernel's timekeeping framework unless explicitly re-engaged by user-level
logic.
Callback functions are provided to perform minimal time accounting when needed, ensuring
correctness of process time tracking while preserving user-space control.

\subsection{Inter-Core Communication for the Isolated Core}

In many real-time and safety-critical applications, the isolated core must exchange information
with other cores to coordinate tasks, report status, or respond to external events. The ability
to communicate efficiently between cores is therefore crucial for both system performance and reliability. 

Currently, the communication between the isolated core and other cores is implemented using
a shared memory mechanism. Shared memory provides a straightforward and low-overhead method for
data exchange, allowing both the isolated core and the other cores to directly read from and
write to common memory regions. Measurements indicate that the round-trip latency of this
communication channel is approximately 300 ns, which is sufficiently low for many real-time
control and monitoring applications (In our evaluation on the Phytium platform---1000 loops,
each with 409{,}600 round trips---the shared memory communication channel achieved a latency
between 200\,ns and 300\,ns).

To ensure data consistency and prevent race conditions, synchronization mechanisms such as
memory barriers, or atomic operations are employed. These mechanisms guarantee that
concurrent accesses to shared memory do not lead to corruption or inconsistent data, which is
particularly important in systems with strict timing and safety requirements.

While the shared memory approach is simple and effective, it may face challenges under scenarios
with high communication frequency or contention. Issues such as cache coherence overhead, access
serialization, or increased latency under load can arise. To address these potential limitations,
future improvements could include the use of hardware-assisted messaging mechanisms.
These enhancements could further reduce latency, increase throughput, and improve the predictability
of inter-core communication.

Overall, the current shared-memory solution provides a practical balance between implementation
simplicity, low latency, and reliability, enabling efficient and deterministic communication
between the isolated core and other cores in the system.

\subsection{GPIO Polling-Based Interrupt Handling for the Isolated Core}

In this approach, during the GPIO polling process, standard interrupt handling disabled to
avoid unnecessary processor interruptions and reduce bus contention. Instead of relying on
conventional interrupt signaling, the system reads the interrupt status registers to determine
whether a GPIO event has been triggered. By directly querying the interrupt controller in this
manner, the processor can detect changes in GPIO levels while minimizing overhead on the system bus.

This method provides several advantages over traditional interrupt-driven mechanisms. First,
by disabling interrupts during polling, unnecessary context switches and interrupt processing
are avoided, which reduces overall system latency. Second, direct access to the interrupt status
registers allows the processor to respond to GPIO events more quickly and predictably. As a result,
this design achieves higher responsiveness and lower bus utilization, which is critical for
real-time and safety-critical applications.

Overall, the combination of GPIO polling with register-based event detection offers a practical
and efficient solution. It balances the need for fast response with reduced system overhead,
enabling the isolated core to handle GPIO-triggered events effectively while maintaining smooth
communication and coordination with other cores.

\section{Evaluation: Deterministic Execution with CPU Isolator}

\label{sec:eval-iso}
We conducted our evaluation of the CPU isolator on a real-time Linux kernel 
derived from the \texttt{linux-6.6-rt} branch of the Phytium Linux Kernel 
repository (version~6.6.63) \cite{phytium-linux-6.6-rt}. Unless otherwise specified, 
all experiments were performed on this kernel with \texttt{PREEMPT\_RT} enabled. 
Importantly, we adapted and extended the kernel to support our isolator, marking 
a key innovation of our work rather than relying solely on the baseline Phytium 
Linux distribution. For completeness, we also maintain support for earlier Linux 
kernels used in legacy deployments; our changes have been kept portable and can 
be applied to older branches with only minor, mechanical adjustments.

The experimental setup consists of the following equipment:  
\begin{itemize}
  \item Signal Generator: UNI-T  
  \item Oscilloscope: Siglent SDS2074X Plus  
  \item Development Board: Phytium Pi E2000Q V2  
  \item Jumper Wires  
\end{itemize}

\begin{figure}
    \centering
    \includegraphics[width=0.85\columnwidth]{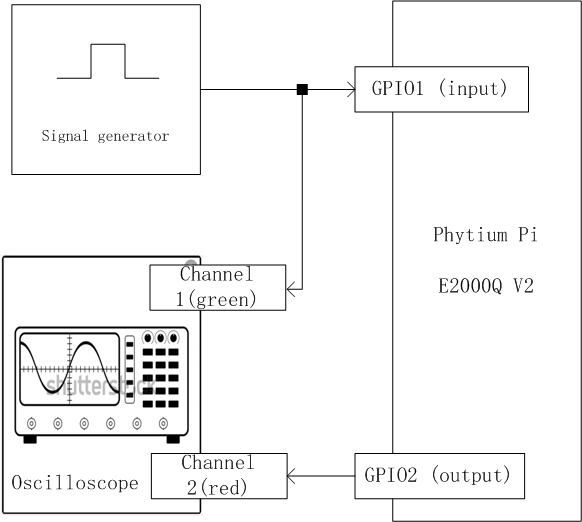}
    \caption{System architecture.}
    \label{fig:oscilloscope}
\end{figure}

The signal generator is connected simultaneously to the GPIO of the Phytium Pi development board and the oscilloscope. The development board, upon detecting the input trigger, produces an output GPIO signal, which is also monitored by the oscilloscope Figure~\ref{fig:oscilloscope}.  

To evaluate determinism, the oscilloscope’s persistence mode is enabled to visualize potential variations in the output signal. The input signal frequency is set to 200~kHz with a voltage range of 0--3.3~V. In order to better capture timing details, the oscilloscope time base is reduced from 2~µs/div to 100~ns/div, thereby improving horizontal resolution. The persistence overlay allows us to observe long-term stability and quantify jitter.

\textbf{Screenshot Analysis:}  
Waveform screenshots are provided, with markers indicating the shortest and longest response times of the output signal relative to the input trigger. The jitter is calculated as the difference between the maximum and minimum response times:  

\[
Jitter = T\_{max} - T\_{min}
\]

Here, the left edge of the persistence trace corresponds to the shortest response, while the right edge corresponds to the longest response.  

\begin{figure}[h]
  \centering
  \includegraphics[width=0.8\linewidth]{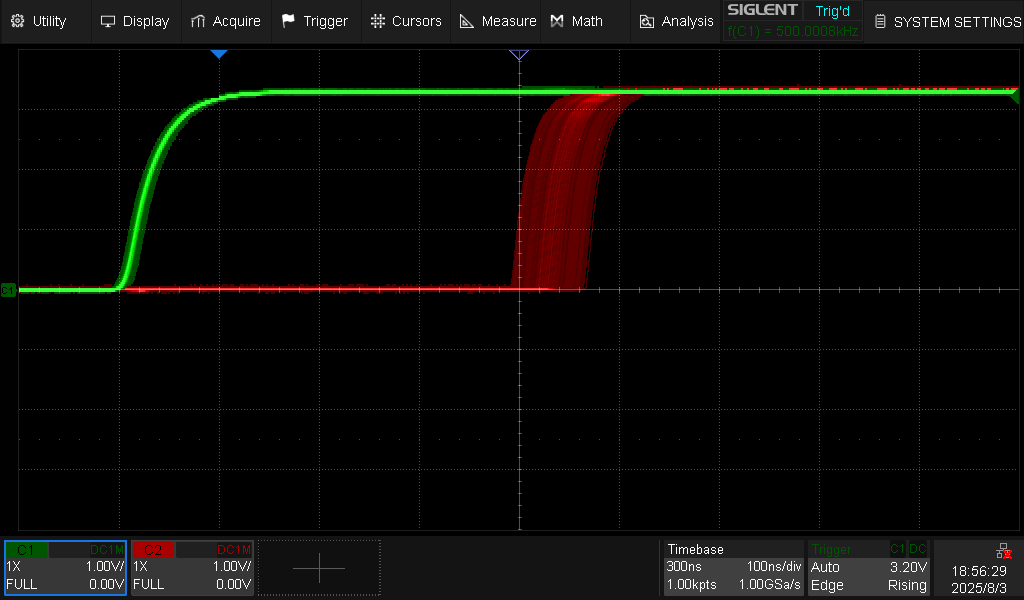}
  \caption{Oscilloscope persistence waveform showing shortest (left) and longest (right) GPIO response times (isolator)}
  \label{fig:islintgpio}
\end{figure}

\subsection{Experimental Results}
In a 24-hour continuous test, the system with isolator technology achieved, as show in Figure~\ref{fig:islintgpio} \footnote {
As shown in Figure~\ref{fig:islintgpio}, the oscilloscope persistence waveform clearly
illustrates the shortest (left) and longest (right) GPIO response times
through the isolator. The green trace consistently departs from the
x-axis at approximately 100~ns, corresponding to the oscilloscope
trigger on channel~1. The red trace departs from the x-axis at times
ranging between about 1 + 3.9 and just a bit more than 1 + 4.6 divisions, consistent with our
reported 390--470~ns range. Measuring from the point at which the trace
leaves the x-axis provides a fair and accurate definition of system
response time: the software cannot begin to respond before channel~1
has left the axis, and channel~2 cannot leave until after the software
has responded. Thus, this measurement captures the full response time,
including GPIO delays, and avoids underestimating latency.
}:
\begin{itemize}
    \item Period: 2 µs
    \item Maximum response latency: 470 ns
    \item Minimum response latency: 390 ns
    \item Jitter: 80 ns
\end{itemize}

\begin{figure}[h]
  \centering
  \includegraphics[width=0.8\linewidth]{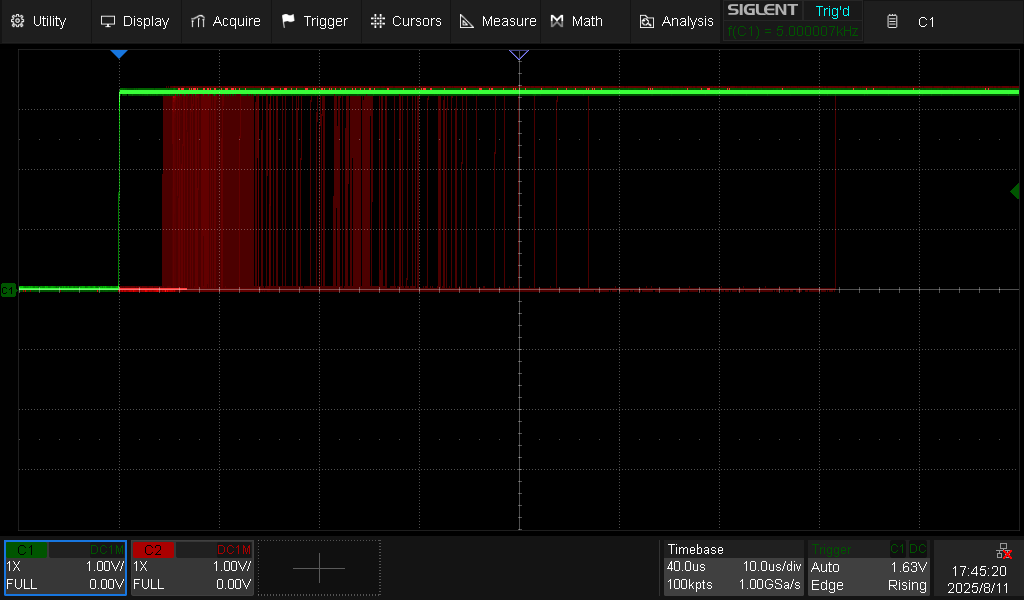}
  \caption{Oscilloscope persistence waveform showing shortest (left) and longest (right) GPIO response times (PREEMPT-RT Linux)}
  \label{fig:linuxintgpio}
\end{figure}

By contrast, in a 24-hour continuous test,  PREEMPT-RT Linux system can only achieve maximum response latency:
72 µs  \footnote {
As shown in Figure~\ref{fig:linuxintgpio}, the oscilloscope persistence waveform clearly
illustrates the shortest (left) and longest (right) GPIO response times
through the isolator. The green trace consistently departs from the
x-axis at approximately 10~µs, corresponding to the oscilloscope
trigger on channel~1. The red trace departs from the x-axis at times
ranging between about 1 + 0.41 and just a bit more than 1 + 7.2 divisions, consistent with our
reported 4100--72000~ns range}.
\begin{itemize}
    \item Period: 200 µs
    \item Maximum response latency: 72  µs
    \item Minimum response latency: 4.1 µs
    \item Jitter: 67.9 µs
\end{itemize}

The results are summarized in Table~\ref{tab:latency-comparison}.

\subsection {Timestamp counter based GPIO toggling test}

\begin{figure}
    \centering
    \includegraphics[width=0.85\columnwidth]{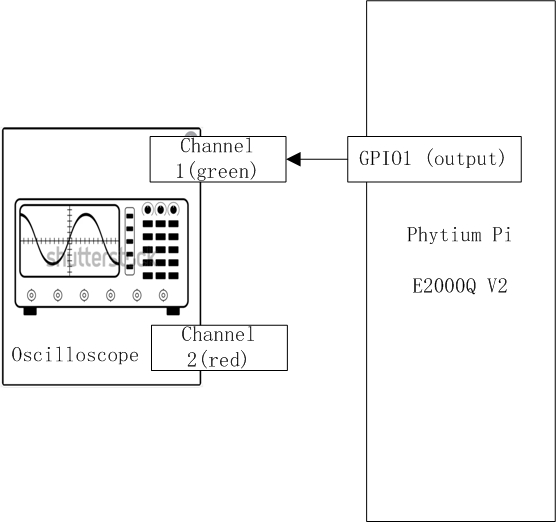}
    \caption{System architecture (Timestamp counter based).}
    \label{fig:oscilloscopets}
\end{figure}
In addition to the external signal generator driven experiment, we implemented a timestamp counter-based GPIO toggling program. In this setup, the output signal is generated periodically by continuously reading the CPU timestamp counter (\texttt{CNTVCT\_EL0}) and flipping the GPIO at predefined intervals, eliminating the need for an external signal source as shown in Figure~\ref{fig:oscilloscopets}.

\begin{figure}[h]
  \centering
  \includegraphics[width=0.8\linewidth]{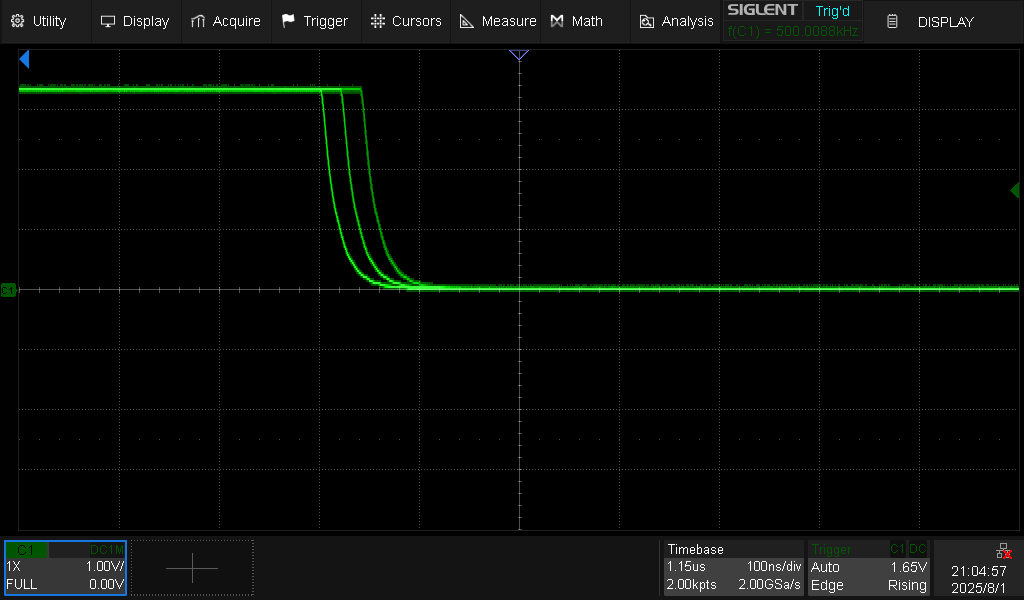}
  \caption{Timestamp counter based oscilloscope maximum deviation (isolator)}
  \label{fig:islcountergpio}
\end{figure}

\begin{figure}[h]
  \centering
  \includegraphics[width=0.8\linewidth]{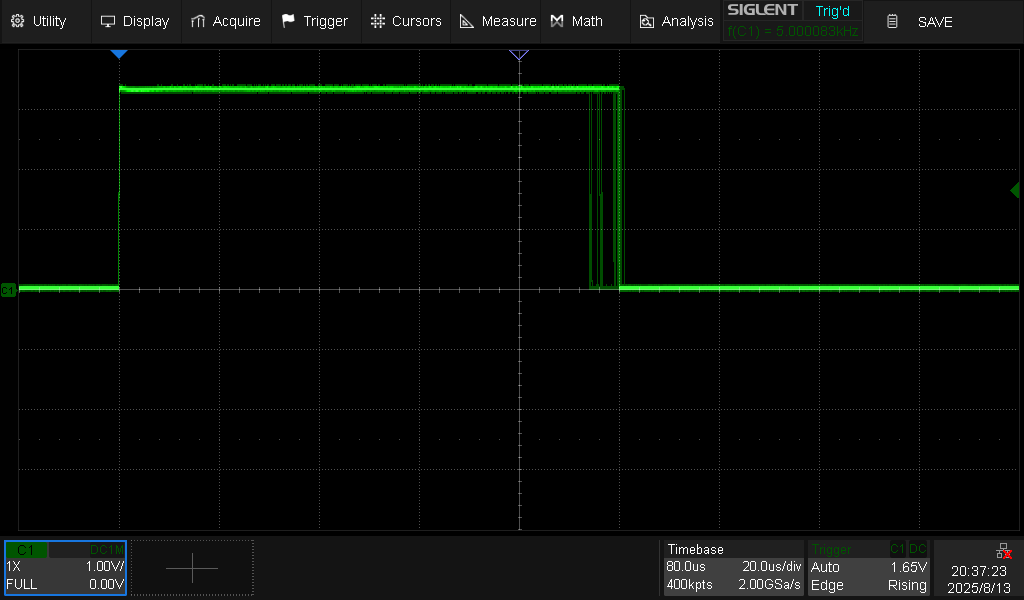}
  \caption{Timestamp counter based oscilloscope maximum deviation (PREEMPT-RT Linux)}
  \label{fig:linuxcountergpio}
\end{figure}

The long-term experiment was conducted for 24 hours under both CPU-isolated and non-isolated environments. The results are summarized in Table~\ref{tab:latency-comparison1}, also showing that the isolator achieves sub-microsecond determinism, while PREEMPT-RT Linux suffers from significantly larger latency
as shown in Figure~\ref{fig:islcountergpio} and Figure~\ref{fig:linuxcountergpio} .

\begin{table}[t]
\centering
\caption{Latency measurements over a 24-hour period (corresponding to Figures~\ref{fig:islintgpio} and~\ref{fig:linuxintgpio}).}
\resizebox{\columnwidth}{!}{%
\begin{tabular}{lccc}
\hline
\textbf{Configuration} & \textbf{Period} & \textbf{Min Lat} & \textbf{Max Lat} \\
\hline
With Isolator     & 2000~ns   & 390~ns   & 470~ns  \\
PREEMPT-RT Linux  & 200000~ns & 4100~ns  & 72000~ns \\
\hline
\end{tabular}
}
\label{tab:latency-comparison}
\end{table}

\begin{table}[t]
\centering
\caption{Latency comparison under different configurations (timestamp counter based), measured over a 24-hour period. Corresponding results are shown in Figures~\ref{fig:islcountergpio} and~\ref{fig:linuxcountergpio}.}
\resizebox{\columnwidth}{!}{%
\begin{tabular}{lccc}
\hline
\textbf{Configuration} & \textbf{Period} & \textbf{Min Lat} & \textbf{Max Lat} \\
\hline
With Isolator     & 2000~ns   & 0~ns   & 40~ns \\
PREEMPT-RT Linux  & 200000~ns & 0~ns   & 8000~ns \\
\hline
\end{tabular}
}
\label{tab:latency-comparison1}
\end{table}

\subsection{Inter-core Communication Delay Measurement}

To further evaluate the determinism of the isolation framework in multicore environments, we
conducted an inter-core communication delay experiment using GPIO and inter-processor interrupts (IPIs).
In this setup, Core~A raises \texttt{GPIO1} and simultaneously sends an IPI to Core~B. Upon receiving
the interrupt, Core~B responds by raising \texttt{GPIO2}.
The other ends of GPIO1 and GPIO2 are connected to Channel 1 and Channel 2 of the oscilloscope respectively
(the same as Figure~\ref{fig:oscilloscope}).
The latency between the two GPIO transitions,
captured by the oscilloscope, directly reflects the inter-core communication delay.

\begin{table}[t]
\centering
\caption{Latency comparison under different configurations (inter-core communication delay), measured over a 24-hour period. Corresponding results are shown in Figures~\ref{fig:islcrossgpio} and~\ref{fig:linuxcrossgpio}.}
\resizebox{\columnwidth}{!}{%
\begin{tabular}{lccc}
\hline
\textbf{Configuration} & \textbf{Period} & \textbf{Min Lat} & \textbf{Max Lat} \\
\hline
With Isolator     & 2000~ns   & 120~ns   & 160~ns \\
PREEMPT-RT Linux  & 200000~ns & 1040~ns  & 2200~ns \\
\hline
\end{tabular}
}
\label{tab:latency-comparison3}
\end{table}

\begin{figure}[H]
  \centering
  \includegraphics[width=0.8\linewidth]{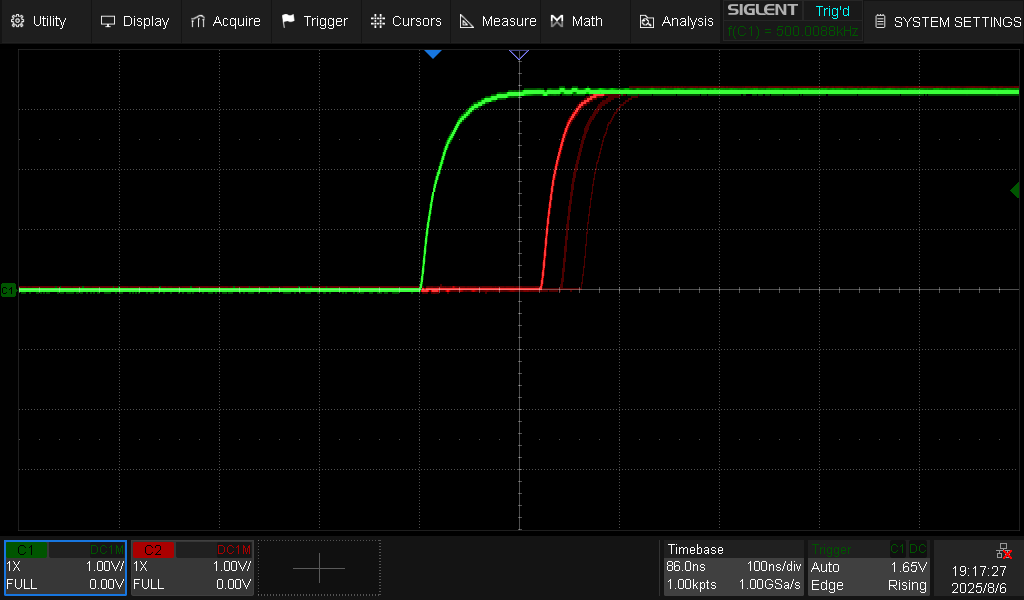}
  \caption{Oscilloscope persistence waveform showing shortest (left) and longest (right) GPIO response times (Isolator cross-core communication).}
  \label{fig:islcrossgpio}
\end{figure}

\begin{figure}[H]
  \centering
  \includegraphics[width=0.8\linewidth]{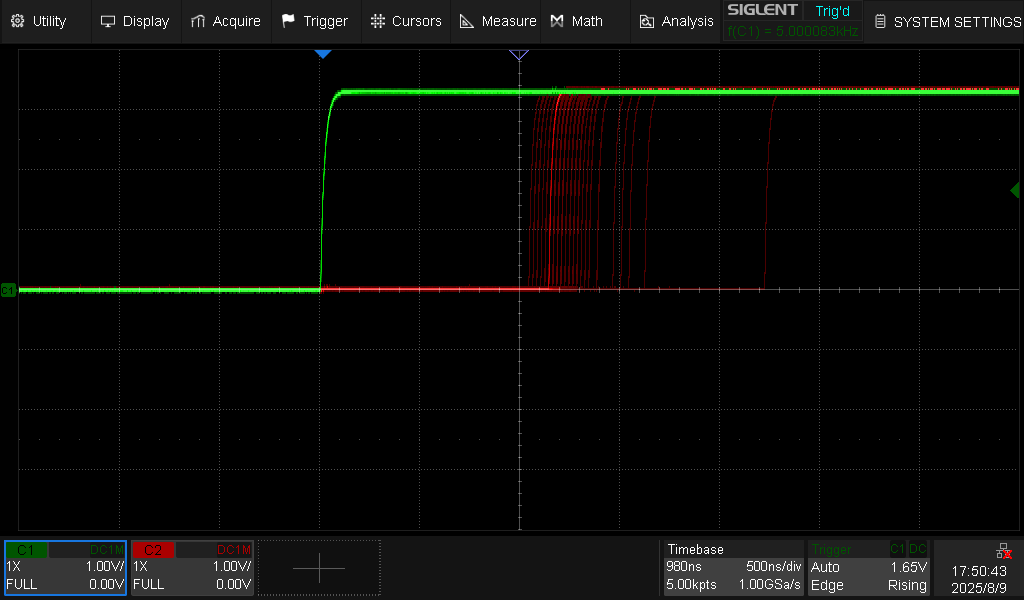}
  \caption{Oscilloscope persistence waveform showing shortest (left) and longest (right) GPIO response times (PREEMPT-RT Linux cross-core communication).}
  \label{fig:linuxcrossgpio}
\end{figure}

Results show that with the Isolator enabled, inter-core communication latency consistently remained
in the sub-microsecond range, with a minimum of about 120~ns, a maximum below 160~ns, and jitter
under 40~ns, as illustrated in Figure~\ref{fig:islcrossgpio}. In contrast, under PREEMPT-RT Linux,
latency exhibited significant variability, with maximum values exceeding 2~µs
(Figure~\ref{fig:linuxcrossgpio}). These findings demonstrate that the isolation framework effectively
eliminates involuntary interruptions and ensures low-jitter, deterministic execution in cross-core
communication scenarios, as summarized in Table~\ref{tab:latency-comparison3}.

\subsection{Oscilloscope Persistence Mode for Long-Term Measurements}

To capture latency characteristics over long observation periods, we employed the persistence
mode of the oscilloscope. In this mode, each newly acquired waveform is overlaid on top of
previous captures. Older traces gradually fade while more recent ones remain bright, resulting
in an accumulated display of thousands or even millions of measurements.

This visualization technique makes it possible to observe both the shortest and longest response
times within a single composite waveform. By continuously running the inter-core communication
test for 24 hours, the persistence display revealed the absolute minimum and maximum GPIO
transition delays, ensuring that rare outliers were not overlooked. As shown in
the Figures in this section, the persistence mode provided
a clear and intuitive representation of latency bounds under both the Isolator and PREEMPT-RT
configurations.

The experiments in this section demonstrate that using the isolator significantly improves GPIO signal determinism on the Phytium Pi development board. When triggered by an external signal generator, the isolator-enabled system achieves sub-microsecond maximum latency and minimal jitter (80 ns), as visualized through the oscilloscope persistence mode. In contrast, a standard PREEMPT-RT Linux system exhibits much higher latency jitter (up to 67.9 µs). The counter-based GPIO toggling test over 24 hours further confirms that CPU isolation with the isolator ensures stable, predictable timing, whereas non-isolated real-time Linux shows large and inconsistent latencies. Overall, the isolator provides robust deterministic execution for time-critical GPIO operations.

\section{Summary}
This paper introduced a kernel-level isolation framework that enables deterministic sub-microsecond execution on Linux by eliminating involuntary interruptions on dedicated cores. The proposed design suppresses periodic tick timers, defers real-time bandwidth enforcement, and replaces asynchronous inter-processor interrupts with shared-memory coordination. Through the Isolator API, applications retain explicit control over when and how deferred maintenance tasks are executed, preserving both determinism and system correctness.

Evaluation on the Phytium Pi platform demonstrates the effectiveness of this approach. Under isolator-enabled operation, GPIO response latencies consistently remain below 470ns with jitter as low as 80ns, compared to up to 72~µs latency with jitter 67.9 µs observed on PREEMPT-RT Linux. A timestamp-counter based GPIO toggling experiment further validated that the isolator achieves stable, sub-microsecond determinism over 24 hours, while non-isolated real-time Linux exhibited latency fluctuations in the hundreds of microseconds.

Together, these results highlight that interrupt isolation and cooperative tick management provide a lightweight yet powerful strategy for achieving predictable performance on general-purpose operating systems. The framework enables strong temporal isolation and low-jitter execution, making it well suited for latency-critical scenarios such as packet processing, edge computing, and real-time control in embedded systems. Notably, the principles outlined in this work are being actively explored in Ucas Wanghuo Linux \cite{zhou2025linux}, where the isolator-inspired mechanisms are integrated into the system architecture to reinforce deterministic execution guarantees. This demonstrates the broader applicability of our design in advancing real-time performance within production-grade operating systems.

\newpage
\bibliographystyle{flairs} 
\bibliography{references.bib}

\begin{thebibliography}{}

\bibitem[\protect\citeauthoryear{Corbet}{2015}]{Corbet}
Corbet, J.
\newblock 2015.
\newblock Dropping the timer tick — for real this time.
\newblock \url{https://lwn.net/Articles/659490/}.

\bibitem[\protect\citeauthoryear{Corbet}{2025}]{KernelDocNOHZ}
Corbet, J.
\newblock 2025.
\newblock No\_hz: Reducing scheduling-clock ticks.
\newblock \url{http://docs.kernel.org/timers/no_hz.html}.

\bibitem[\protect\citeauthoryear{Deng \bgroup et al\mbox.\egroup
  }{2024}]{YuxinRen}
Deng, Z.; Zhang, Z.; Li, D.; Guo, Y.; Ye, Y.; Ren, Y.; Jia, N.; and Hu, X.
\newblock 2024.
\newblock A 6 years’ experience in mitigating cross-core interference in
  linux.
\newblock In {\em 2024 IEEE Real-Time Systems Symposium (RTSS)}.
\newblock 2024 IEEE Real-Time Systems Symposium (RTSS).

\bibitem[\protect\citeauthoryear{{Linux kernel
  community}}{2025}]{KernelDocParam}
{Linux kernel community}.
\newblock 2025.
\newblock The kernel’s command-line parameters.
\newblock \url{http://docs.kernel.org/admin-guide/kernel-parameters.html}.

\bibitem[\protect\citeauthoryear{Mellichamp}{1983}]{mellichamp1983}
Mellichamp, D.~A.
\newblock 1983.
\newblock {\em Real-time computing: with applications to data acquisition and
  control}.
\newblock New York: Van Nostrand Reinhold.

\bibitem[\protect\citeauthoryear{{Phytium Linux}}{}]{phytium-linux-6.6-rt}
{Phytium Linux}.
\newblock Phytium linux kernel: \texttt{linux-6.6-rt} branch.
\newblock \url{https://gitee.com/phytium_embedded/phytium-linux-kernel}.
\newblock Kernel version 6.6.63.

\bibitem[\protect\citeauthoryear{{Redhawk Linux}}{2025}]{redhawk}
{Redhawk Linux}.
\newblock 2025.
\newblock Redhawk linux rtos.
\newblock \url{https://concurrent-rt.com/products/software/redhawk-linux/}.

\bibitem[\protect\citeauthoryear{{Risk Magazine}}{2024}]{risk2024ultra}
{Risk Magazine}.
\newblock 2024.
\newblock Ultra-low latency trading: how low can you go?
\newblock
  https://www.risk.net/insight/technology-and-data/7960459/ultra-low-latency-trading-how-low-can-you-go?
\newblock Accessed: 2025-09-08.

\bibitem[\protect\citeauthoryear{stackoverflow}{2020}]{stackoverflow}
stackoverflow.
\newblock 2020.
\newblock Completely eliminating the timer tick in modern linux >=5.0.
\newblock
  \url{https://stackoverflow.com/questions/60322119/completely-eliminating-the-timer-tick-in-modern-linux-5-0}.

\bibitem[\protect\citeauthoryear{Torvalds}{2025}]{torvalds2025linux}
Torvalds, L.
\newblock 2025.
\newblock Merge tag 'for\_linus' of
  git://git.kernel.org/pub/scm/linux/kernel/git/mst/vhost.
\newblock Accessed: 2025-08-27.
\newblock Commit ID: 39f90c1, Merge: 518b21b 45d8ef6.

\bibitem[\protect\citeauthoryear{Vahalia}{1996}]{Vahalia96}
Vahalia, U.
\newblock 1996.
\newblock {\em {UNIX} Internals: The New Frontiers}.
\newblock Prentice Hall.

\bibitem[\protect\citeauthoryear{Weisbecker}{2012}]{weisbecker2012context}
Weisbecker, F.
\newblock 2012.
\newblock context\_tracking: New context tracking subsystem.
\newblock
  \url{https://git.kernel.org/pub/scm/linux/kernel/git/torvalds/linux.git/commit/?id=91d1aa43d30505b0b825db8898ffc80a8eca96c7}.
\newblock Commit 91d1aa43d30505b0b825db8898ffc80a8eca96c7 in the Linux Kernel
  repository.

\bibitem[\protect\citeauthoryear{Zhou \bgroup et al\mbox.\egroup
  }{2025}]{zhou2025linux}
Zhou, Z.; Liu, Z.; Zhang, S.; Li, J.; Du, D.; Sun, M.; Wang, Z.; Liu, H.; and
  Xu, G.
\newblock 2025.
\newblock Linux extreme exploration {II}: Achieving sub-microsecond system
  response with real-time linux (online video in chinese.
\newblock
  \url{https://www.bilibili.com/video/BV1KZaczLECN/?share_source=copy_web&vd_source=c469f17afc4323fbccad8d3906d31d25}.
\newblock Online Video in Chinese; accessed 3 September 2025.

\end{thebibliography}

\end{document}